\documentclass[useAMS,usenatbib,usegraphicx]{mn2e}
\title[Non-radial pulsations in $\gamma$\,Bootis ]
  {A spectroscopic search for non-radial pulsations in the $\delta$\,Scuti star
$\bgamma$\,Bootis}
\author[R. Ventura et al.]
  {R. Ventura$^1$, G.~Catanzaro$^1$, J.~Christensen-Dalsgaard$^2$,
  M.P.~Di Mauro$^3$, 
 \newauthor
and L.~Patern\`o$^4$\\
  $^1$INAF - Osservatorio Astrofisico di Catania, Via S. Sofia 78, 95123 Catania, Italy\\
$^2$Institut for Fysik og
                Astronomi, Bygn. 1520, Aarhus Universitet, Ny Munkegade,
              DK-8000 Aarhus C, Denmark\\
  $^3$INAF - IASF, Istituto di Astrofisica Spaziale e Fisica Cosmica, Via del Fosso del Cavaliere 100, 00133 Roma, Italy\\
  $^4$Dipartimento di Fisica ed Astronomia, Universit\`a di Catania, Via S. Sofia 78, 95123 Catania, Italy}

\date{Accepted 2007 August 09. Received 2007 August 09; in original form 2007 June 22}

\pagerange{\pageref{firstpage}--\pageref{lastpage}} \pubyear{2006}

\def\LaTeX{L\kern-.36em\raise.3ex\hbox{a}\kern-.15em
    T\kern-.1667em\lower.7ex\hbox{E}\kern-.125emX}

\begin{document}

\label{firstpage}

\maketitle

\begin{abstract}
High-resolution spectroscopic observations of the rapidly rotating $\delta$\,Scuti star $\bgamma$\,Bootis have 
been carried out on 2005, over 6 consecutive nights, in order to search for line-profile variability.
Time series, consisting of flux measurements at each wavelength bin across the Ti{\sc ii} 4571.917{\AA} 
line profile as a function of time, have been Fourier analyzed.

The results confirm the early detection reported by \citet{kennelly92} of a dominant periodic component 
at frequency $21.28\,{\rm~c/d}$ in the observer's frame, probably due to a high azimuthal order sectorial mode. 
Moreover, we found other periodicities at 5.06~c/d, 12.09~c/d, probably present but not secure, and at
11.70~c/d and 18.09~c/d, uncertain. The latter frequency, if present, should be identifiable as another 
high azimuthal order sectorial mode and three additional terms, probably due to low-$l$ modes, as proved by the 
analysis of the first three moments of the line.
Owing to the short time baseline and the one-site temporal sampling we consider our results
only preliminary but encouraging for a more extensive multisite campaign.

A refinement of the atmospheric physical parameters of the star has been obtained from our spectroscopic data 
and adopted for preliminary computations of evolutionary models of $\bgamma$\,Bootis.

\end{abstract}


\section{Introduction}
$\delta$\,Scuti stars are typically population I, A and F main-sequence or slightly post-main-sequence 
variables of mass $M= 1.4 - 3.0 {\rm M}_{\odot}$, either in a stage of H-core or H-shell burning, with projected 
rotational velocity $v \sin i$ in the range of $10-200\,{\rm km/s}$.
They are located in the lower part of the classical instability strip where the $\kappa$ mechanism 
is expected to drive radial and non-radial p modes,
modes with mixed p and g character and possibly also g modes.
Many of these variables are observed to
be low-amplitude, multi-mode pulsators 
with power spectra sufficiently rich to allow an asteroseismological
investigation.

Photometry of $\delta$\,Scuti stars has allowed to identify low-degree ($l \leq 3$), low-order ($n \leq 7$) 
p modes (see \citealt{garrido00} and \citealt{poretti00} for exhaustive
reviews), while additional information came from spectroscopic studies which have confirmed the
presence of high-degree non-radial modes with $l$ up to $20$ (\citealt{mantegazza00}) in moderate or fast rotators.

In these stars pulsation modes of high degree manifest themselves as cyclical line-profile variability characterized 
by the appearance of weak (less than 1\% of continuum) absorption moving bumps travelling across the rotationally 
broadened profiles from side to side.
Rapid rotation enhances the visibility of high-degree modes because it allows to spatially resolve the wave 
pattern of the photospheric velocity fields
across the stellar disk, leading to a one-to-one correspondence between wavelength in the broaded 
line profile and position on the stellar surface (the so-called Doppler imaging). Adjacent velocity fields of 
opposite sign, traveling around the star due to rotation, produce opposite Doppler shifts, effectively causing 
a re-distribution of the flux across the rotationally broadened line profile and then the appearance of 
characteristic absorption moving distortions (\citealt{vogt83}). The apparent period of variations is 
the time for the pattern to repeat itself and it is related to both the phase velocity of the wave and the 
rotational velocity of the star.

\begin{table}
\caption{Observation log of $\gamma$\,Bootis spectroscopy}
\label{log}
\begin{center}
\begin{tabular} {@{}ccc}
\hline
\hline
Heliocentric Julian Date & $\delta T$ & Number of spectra \\
(2450000+) & (hours)& \\
3543.31958 & 6.28 &  67 \\
3544.37164 & 3.77 &  43 \\
3545.31669 & 6.44 &  71 \\
3546.33466 & 5.56 &  59 \\
3547.31994 & 5.83 &  65 \\
3548.31679 & 5.62 &  62 \\
\hline
\end{tabular}
\end{center}
\end{table}

Moving bumps in $\delta$ Scuti stars were first observed by \citet{yang86} and \citet{walker87}. 
From then on, the number of rapidly rotating $\delta$\,Scuti stars showing high-degree modes has increased 
progressively thanks both to the improvements in spectroscopic techniques and theoretical developments which 
have triggered an extensive monitoring of stars suspected to be non-radial pulsators. To date, 16 $\delta$\,Scuti 
high-degree pulsators (see \citealt{mantegazza00}, \citealt{balona01} and \citealt{koen02}) have been studied in 
detail for line-profile variability.

Among these stars, $\gamma$\,Bootis (HD127762, A7III) is a bright ($V = 3$), rapidly rotating $\delta$\,Scuti star 
that belongs to a transition group located between the main-sequence variables and the Cepheids. Photometric 
variations with a period of $P = 0.25$ days have been reported by \citet{auvergne79}. 
This star is among the few $\delta$ Scuti stars with a period longer than 0.20 days 
(see \citealt{rodriguez00}).
Line-profile variations in the spectra of the star were detected for the first 
time by \citet{kennelly92}. The authors collected a total of 29 spectrograms, at a time cadence of about 10 min, 
in two different periods, three months apart, covering a total observing time of about 5 hours. They analyzed the 
data by sampling at 1.5\,{\AA} intervals the intensity of the residuals built by subtracting the mean profile from 
each profile in the series, and then computing the amplitude spectra in the Fourier domain. A period of variation 
in moving distortions of
$0.047 \pm 0.002$ days was found and interpreted as a high-degree ($\vert m\vert \simeq 10$) non-radial 
mode with a period of $0.073$ days in the corotating frame. The series were not long enough to cover a full 
cycle of the known photometric period reported by \citet{auvergne79} so that \citet{kennelly92} obtained only 
a rough confirmation of it. Since then, to the best of our knowledge, no further and more accurate investigations 
on pulsations of the star have been reported in literature.

In this paper we present the results of new careful spectroscopic observations of $\gamma$\,Boo performed 
on a baseline of 6 nights at the {\it INAF - Catania Astrophysical Observatory} and compare them with 
theoretical predictions deduced from a grid of models of the star computed by including the effect of rotation 
on the stellar model structure and on 
global oscillations.
In Section~2 we describe the acquisition and reduction of the data. In Section~3 we discuss the procedures 
adopted to determine the atmospheric parameters of the star, necessary to construct realistic models of the star, 
and those to analyze the line-profile variability and present the results.
Finally, in Section~4 we discuss the implications of our finding and draw some conclusion.
\begin{table}
\caption{Parameters of $\gamma$\,Bootis}
\label{par}
\begin{center}
\begin{tabular} {@{}ll}
\hline
\hline

 Parallax      &  38.29 $\pm$ 0.73 mas (Hipparcos Main Catalogue) \\
 V             & 3.04 \\
 B-V           & 0.191 \\
 U-B           & 0.120 \\
 Spectral Type & A7 III\\
 $v \sin i$        & 145 km/s (\citealt{auvergne79})\\
               &  123 km/s  (\citealt{erspamer03})\\
               &  128 km/s  (\citealt{royer02})\\
               &  115 km/s (\citealt{abt95})\\
  $T_{\rm eff}$    &  8000 K (\citealt{auvergne79})\\
               &  7585 K (\citealt{erspamer03})\\
  $M_V$  & 0.93 \\
  Photometric frequency & 0.25 d (\citealt{auvergne79})\\
  Spectroscopic frequency & 0.0047 d (\citealt{kennelly92})\\
\hline
\end{tabular}
\end{center}
\end{table}

\section{Observations and data reduction}
Time-resolved spectroscopy of $\gamma$\,Boo was carried out during six
consecutive nights, from June 21 to June 26 2005, by means of the REOSC echelle
spectrograph operating with the $91\,{\rm cm}$ telescope of the {\it INAF - Catania Astrophysical
Observatory}.

The spectrograph is fiber-linked to the telescope
through an UV-NIR type fiber of $100\,\mu$m core
diameter. It is designed to work both at high resolution, in cross-dispersion mode, and at low
resolution, in single-dispersion mode. The observations were performed in the  cross-dispersion 
con\-fi\-gu\-ra\-tion making use of both gratings: the 300\,grooves/mm echellette grating, blazed at 4.3 deg 
whose maximum efficiency is about 80$\%$ at the blaze wavelength 5000 {\AA}, and the echelle
grating with $79\, {\rm grooves/mm}$ blazed at $63.443\,{\rm deg}$. Spectra were acquired through a
thinned, back-illuminated (SITE) CCD with 1024 x 1024 pixels of $24\,\mu$m
size, whose typical readout noise is about 6.5 e$^{-}$ and photon gain $2.5\,\rm{ph/ADU}$. The resulting
spectral range was approximately from $4300${\AA} to $6700${\AA} (19 orders).

A total of 367 spectrograms, covering $33.5\,{\rm hours}$ of observation (see Table~\ref{log}), were obtained.  
The spectrograms were extracted from the CCD
images and calibrated in wavelengths by using the NOAO/IRAF package.
In the spectral region covered by our observations, owing to the relatively high value of the
projected rotational velocity, only the Ti{\sc ii} 4571.917{\AA} line was found completely
free from blends of adjacent features, thus suitable to allow a good normalization to the stellar
con\-ti\-nu\-um and a study of the line-profile variability.
In the spectral region of the selected line the resulting reciprocal dispersion was
0.12 px mm$^{-1}$ with an effective resolving power of R\,$\approx$\,16000, as deduced from
the emission lines of the Th-Ar calibration lamp. The integration time for each exposure was set to
5 minutes with a resulting signal-to-noise ratio up to 200. This exposure time was optimized in such a way 
as to avoid phase-smearing effects.
Finally, the spectra have been corrected in order to remove the observer's velocity variation due to the Earth's 
revolution and rotation, and the
con\-ti\-nu\-um in the proximity of the line was defined by a linear least-squares
fit of two spectral windows selected on both red and blue sides of the line.

\section{Data processing}
\subsection{Determination of the atmospheric parameters and abundance analysis}

\begin{figure}
\includegraphics[height=9cm,angle=90]{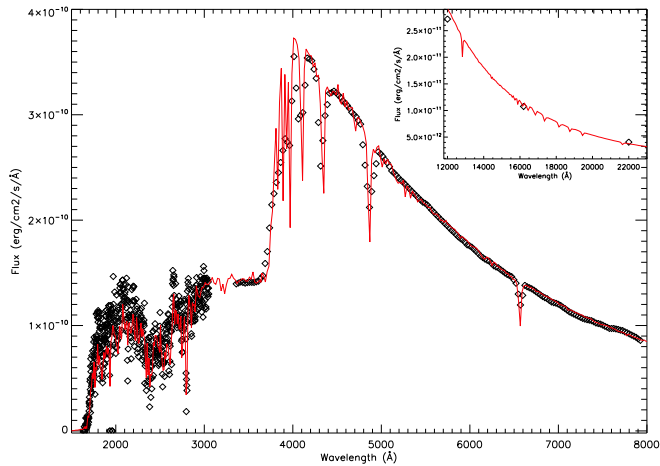}
\caption[]{Observed spectral energy distribution (SED) built from various
literature sources as described in the text. Thick line represents theoretical SED computed by 
ATLAS9 code for $T_{\rm eff}=7600\,{\rm K}$ and $\log g=3.70$.}
\label{sed}
\end{figure}

The atmospheric structure of $\gamma$\,Bootis has been studied by several authors
in the last decades (see Table~\ref{par}). \citet{auvergne79} found an effective temperature $T_{\rm eff}=8000\, {\rm K}$,
a surface gravity $\log g = 3.8$ and a projected rotational velocity $v \sin i=145\,{\rm km/s}$, values adopted later by
\citet{kennelly92}.

More recently, \citet {erspamer03}, based on Elodie spectrograms,
derived for the star a slightly lower effective temperature $T_{\rm eff}=7585\,{\rm K}$
and values of surface gravity $\log g=3.75$, microturbulent velocity
$\xi=3.1\,\rm{km/s}$ and $v \sin i=123\,{\rm km/s}$. The last value is also
in good agreement with that reported by  \citet{royer02}, based on Aurelie spectrograms,
who found $v \sin i= 128\,{\rm km/s}$, while \citet{abt95}
derived from CCD Coud\'e  spectra a slightly lower value of $115\,{\rm km/s}$.

To determine effective temperature and surface gravity for our star, we proceeded in
two consecutive steps: (i) we derived atmospheric parameters by matching the observed flux
distribution obtained by joining various data sources collected from the literature and
(ii) we refined the values found in the previous step by modelling the observed
H$_\beta$ line profile.\\
In step (i), the observed flux (see Fig.~\ref{sed}) has been built using the following data:
\begin{itemize}
\item IUE spectra processed with the NEWSIPS reduction method taken from
      {\it INES} Final Archive data. The data consist of two high-dispersion spectra
      (SWP44484RL and LWP25820RL), that we co-added manually for obtaining a unique spectrum
      covering the 1150-3350 {\AA} interval;
\item spectrophotometry in the range 3200--7900 {\AA} taken from \citet{burnashev85};
\item {\it UBV} magnitudes from \citet{john53};
\item {\it JHK} magnitudes from {\it 2MASS} survey \citep{skrut03}.
\end{itemize}
The resulting spectrum has been de-reddened following the procedure by \citet{cardelli89}
and then compared to a grid of theoretical fluxes computed by means of ATLAS9 code
\citep{Kur93} with convection turned on ($l/H = 1.25$). The colour excess, E(B-V), needed by the de-reddening 
procedure has been derived from E(b-y) = 0.01 \citep{gray01} using the relation
E(b-y) = 0.74 E(B-V) \citep{crawford75}. The starting values of
$T_{\rm eff}$ and $\log g$ have been obtained from the Str\"omgren photometry of $\gamma$\,Bootis \citep{hauck98} 
according to the grid of \citet{md85}. The photometric colours have been de-reddened with the \citet{moon85} algorithm. 
From this calculation we found $T_{\rm eff}=7600\,{\rm K}$
and $\log g=3.70$.

\begin{figure}
\includegraphics[width=9cm]{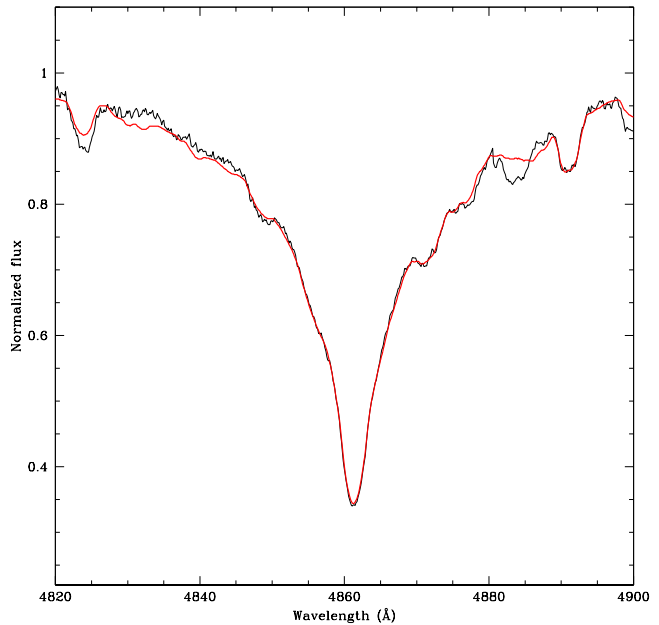}
\caption[]{Comparison between the observed and computed H$_\beta$ line
profile.}
\label{hbeta}
\end{figure}

In step (ii), we compared the observed profile of the H$_\beta$ line
with the synthetic one, minimizing the  differences between them by using the $\chi^2$-criterion to
evaluate the goodness-of-fit. Errors were estimated
as the variation in the parameters which increases the $\chi^2$ by a unit.
The observed spectrum we use from here on is the average of all the collected data.
The final S/N ranges between $230$ and $360$.

By adopting the iterative procedure described in \citet{cat04}, which takes into account
also metal abundances, and by using as input data the values derived in step (i), we obtained the 
final values: $T_{\rm eff} = 7600 \pm150 \,\rm{K}$ and
$\log g = 3.75\pm 0.05$, which are in good agreement with those obtained by
\citet{erspamer03}. 
With these values we then computed the synthetic profile shown in Fig.~\ref{hbeta} by
means of the SYNTHE \citep{Kur81} spectral synthesis code for a LTE atmospheric model with
solar metallicity (Opacity Distribution Function ODF = [0.0]) calculated by using the
ATLAS9 \citep{Kur93} stellar atmosphere program.

The projected rotational velocity has been previously estimated by a non-linear least-squares fit of a
rotationally broadened Gaussian profile to the average line profile of the Ti{\sc ii}
$\lambda$4571.917{\AA}.
The derived value of $115\pm2\,{\rm km/s}$ is in excellent agreement
with those previously derived by \citet{royer02} and \citet{abt95}.
The calculated values of $T_{\rm eff}$, $\log g$, $v \sin i$ are listed in Table~\ref{parnos}.
\begin{table}
\caption{Basic parameters of $\gamma$\,Bootis as derived by the present analysis}
\label{parnos}
\begin{center}
\begin{tabular} {@{}ll}
\hline
\hline
 $v \sin i$        & $115 \pm 2$ km/s \\
  $T_{\rm eff}$    &  $7600\pm150\,\rm{K}$ \\
  $\log g$ & $3.75\pm0.05$ \\
\hline
\end{tabular}
\end{center}
\end{table}

In order to calculate the atmospheric element abundances of $\gamma$ Boo we
computed the synthetic spectrum of the star in the whole observed
spectral region by assuming
the above derived values of $T_{\rm eff}$, $\log g$, $v \sin i$
and the
microturbulent velocity $\xi$ by \citet{erspamer03}.
We used ATLAS9 \citep{Kur93} in order to
compute the LTE atmospheric model and SYNTHE \citep{Kur81} program in order
to identify the observed spectral lines and derive its chemical abundances. The LINUX version of
both codes are those implemented by
\citet{sbordone04}. The values of the oscillator strengths $\log gf$ for each spectral line
are those published by \citet{kur95} and the subsequent upgrading by \citet{castelli04}.
The derived abundances in terms of solar abundances 
$[{\rm X/H}]=\log(N_{\rm X}/N_{\rm H})_\star-\log(N_{\rm X}/N_{\rm H})_\odot$, assuming the solar values by
\citet{grevesse98}, are reported in Table~\ref{abund}.
 For comparison we report also the abundances derived by
\citet{erspamer03}. The deduced uncertainties are typically of the order of 0.1 dex.

As a general result we obtained for $\gamma$ Bootis chemical abundances quite
close to the solar values at least up to Ni (Z=28), a common feature in the
abundance pattern of many $\delta$\,Scuti stars \citep{yush05}.
\citet{erspamer03} stated that in the range of
$T_{\rm eff}$
and $\log g$ they have considered, errors on abundances are typically
less than 0.15~dex.
This results allow us to conclude that abundances derived in our study,
and reported
in Table~\ref{abund}, are compatible with their values, the discrepancies being
always consistent with experimental errors.

\subsection{Line-profile variations}
As mentioned in Section~2, owing to the relatively high rotational velocity of the star, the only spectral
line completely free from line-blend contamination, and therefore suitable for a study of line-profile variability, 
was the Ti{\sc ii} 4571.917{\AA} line.

The Fourier analysis of the Ti{\sc ii} 4571.917{\AA} line-profile variations was performed
by adopting the so-called pixel-by-pixel method (\citealt{mantegazza00}), based on the fact that 
the flux measured at each wavelength bin across a line profile fluctuates with the same period as a 
wave propagating in the photosphere of the star. Therefore from the whole set of 367 residual spectrograms  
we extracted 40 time series (as many as the pixels sampling the Ti{\sc ii} 4571.917{\AA} line profile) consisting 
of the measured fluxes at each wavelength across the line profile as a function of time. The Fourier transform 
of the resulting time strings was performed by adopting the generalized version of the least-squares multiple 
sinusoid fit with known constituents (\citealt{mantegazza95}), originally developed by \citet{vanicek71}. 
A ``global" least-squares spectrum, which contains the contribution to the variability coming -- on the whole --
from the full set of the time series (i.e. from the whole line profile), is computed iteratively
every time a ``known periodic constituent" has already been detected and a new $n+1$ periodic term 
is searched for. A simultaneous least-squares fit of $n+1$ sinusoids is performed exploring the entire 
domain of trial frequencies $\nu_i$,  and a power spectrum is generated by defining the
global reduction factor:
$$ RF_i = 1 - \sum_{j,k} w^2_k(p_{i,j}(t_k) - P(j,t_k))^2/\sigma_n$$
where $p_{i,j}(t_k)$ is the multi-sinusoidal function adopted to fit the $j$-th time series, $P(j,t_k)$, 
sampled at time $t_k$, $\sigma_n$ is the global residual variance after the fit of the line-profile variations 
with the $n$ known constituents, and $w_k$  are the normalized weights derived from the S/N ratio of the 
spectrograms. The frequency $\nu_i$ corresponding to the highest value of {\it RF\/} is then selected 
as the $(n+1)$-th known constituent
and the procedure is iterated again until when no dominant peaks appear in the last power spectrum. 
The technique does not rely on pre-whitening of the data, and amplitudes and phases of the known constituents 
are recomputed, together with their formal errors, each time a new periodic component has been detected.  
A more detailed description of this approach can be found in \citet{mantegazza99} and in \citet{mantegazza00}.
\begin{table}
\label{abu}
\caption{Abundances of chemical elements derived for $\gamma$\,Bootis with
respect to their abundances in the solar atmosphere \citep{grevesse98} expressed as 
$[{\rm X/H}]=\log(N_{\rm X}/N_{\rm H})_\star-\log(N_{\rm X}/N_{\rm H})_\odot$. For
comparison we report the abundances determinated for $\gamma$ Bootis by \citet{erspamer03}.}
\label{abund}
\begin{center}
\begin{tabular} {@{}l|cc}
\hline
\hline
El &  This study & E\&N (2003) \\
   &             &  \\
\hline
Na & ~~0.45  & ~~0.25  \\
Mg & $-$0.05 & ~~0.16  \\
Si & $-$0.01 & ~~0.06  \\
Ca & $-$0.01 & ~~0.04  \\
Sc & $-$0.04 & ~~0.35  \\
Ti & ~~0.25  & ~~0.01  \\
Cr & $-$0.17 & $-$0.14  \\
Mn & $-$0.01 & $-$0.20  \\
Fe & $-$0.08 & $-$0.20  \\
Ni & $-$0.22 & $-$0.22  \\
Ba & $-$0.10 & ~~0.15  \\
\hline
\end{tabular}
\end{center}
\end{table}
On adopting the procedure described above, we computed the least-squares global power spectrum 
by scanning the whole frequency range from 0 up to the Nyquist frequency, which is around 140~c/d, with 
a spacing of 0.04~c/d. However, all the power spectra reported here have been truncated at 40~c/d since all 
the significant power appears at frequency well below 30~c/d. The total length of the data set is such that the 
HWHM of the main power peak in the window function is 0.095~c/d.
\begin{figure*}
\includegraphics[width=18cm]{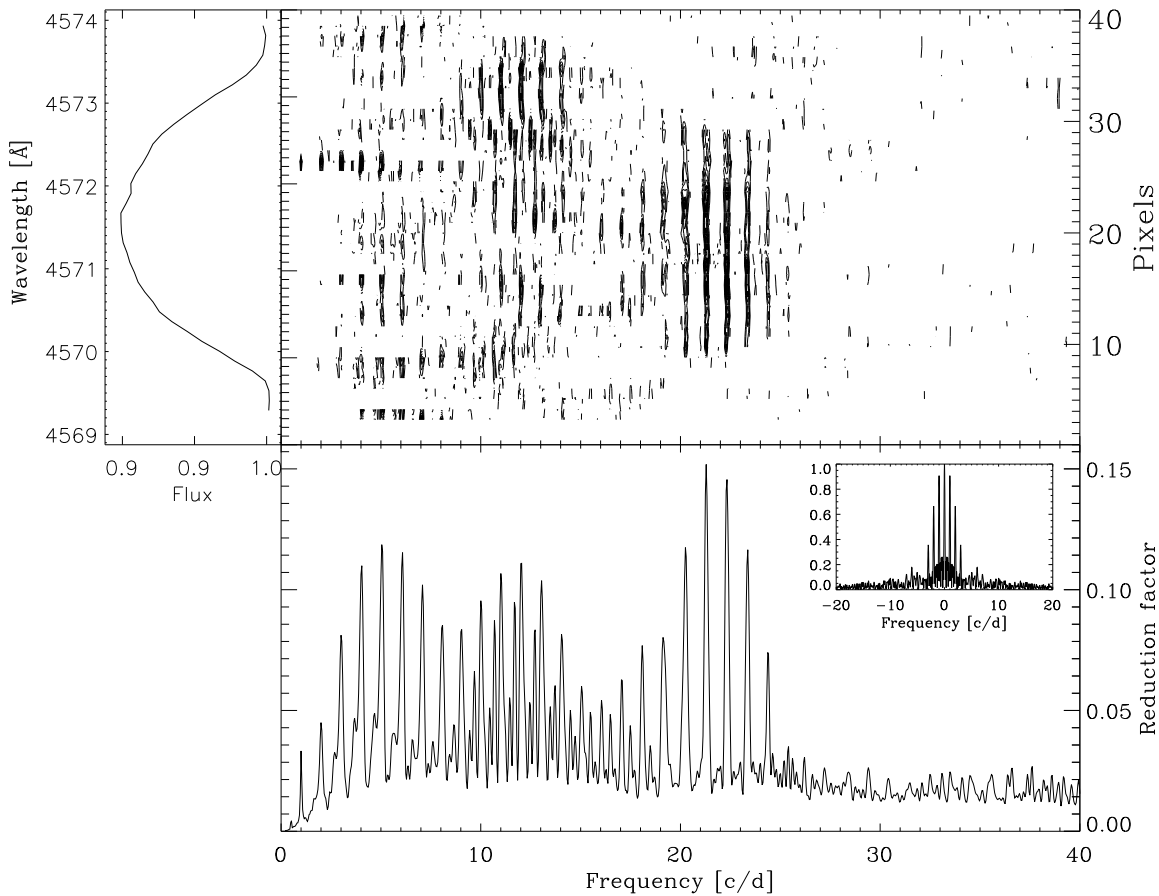}
\caption[]{Pixel-by-pixel least-squares analysis of the variability of the Ti{\sc ii} 4571.917{\AA} line profile. 
Top-left panel: average profile. Top-right panel: pixel-by-pixel least-squares power spectra without known constituents, 
describing the evolution of the
Ti{\sc ii} 4571.917{\AA} line-profile variability as a function of the wavelength bin across the line profile. 
Bottom panel: global least-squares
power spectrum without known constituents. The inset shows the amplitude spectrum of the window function.}
\label{figspettro}
\end{figure*}
\begin{figure*}
\includegraphics[width=18cm]{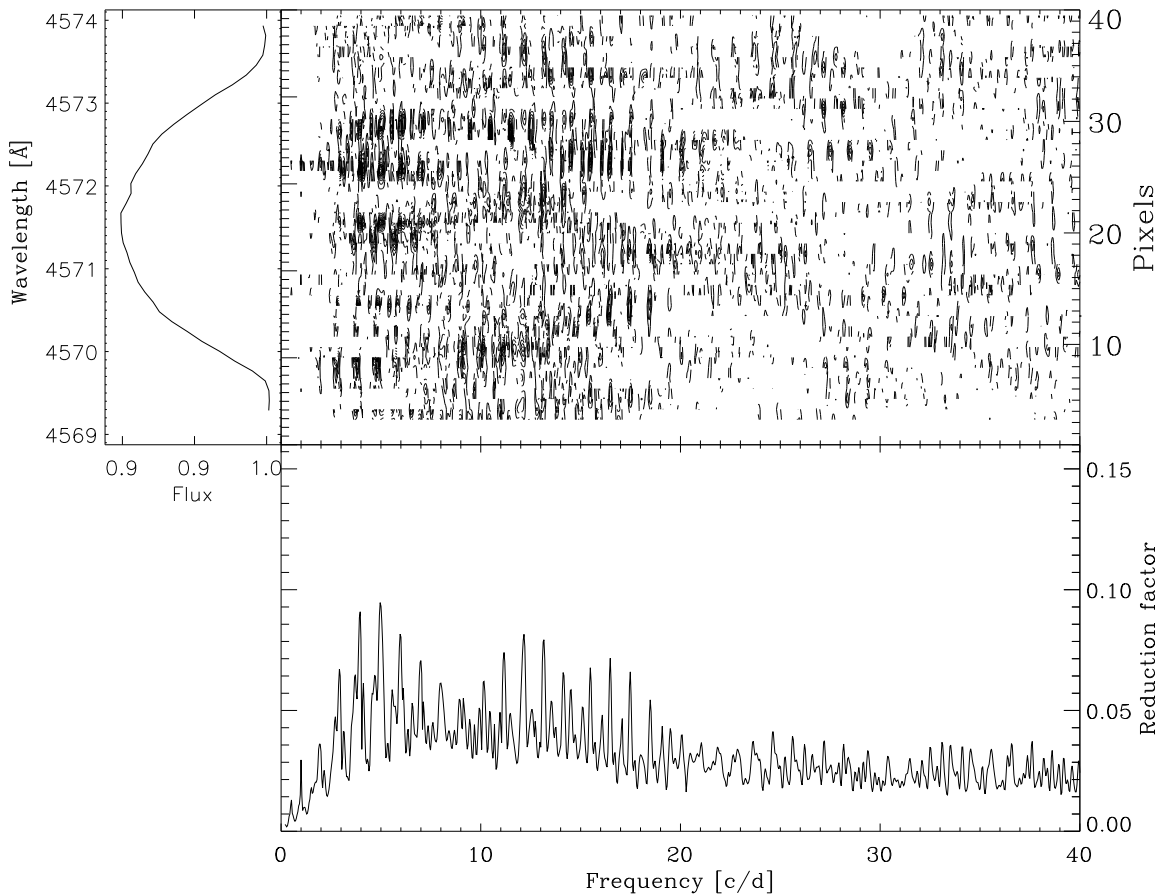}
\caption[]{Pixel-by-pixel least-squares power spectra and global least-squares power spectrum obtained after 
all the known constituents have been removed from the data. In the upper panel the noise  appears to be 
increased with respect to Fig.~\ref{figspettro} because each panel has been normalized  to its maximum value.
}
\label{figspettrof}
\end{figure*}

Fig.~\ref{figspettro} reports both the least-squares pixel-by-pixel power spectra without known constituents, 
which describes the evolution of the variability across the line profile, and the corresponding global spectrum. 
It is evident in both of them  the presence of a
prominent periodicity centred at 21.28~c/d, also reported by \citet{kennelly92} in their 
analysis of line-profile variability. A very sharp thread of power corresponding to this frequency 
extends all over the core of the line profile and beyond, and covers about the 55\% of the whole profile.
Quite strong daily aliases of the central peak, produced by the limited single-site temporal 
sampling (see the spectral window in the inset of the figure) distributed over 6 consecutive nights, 
flank the true frequency. Significant concentrations of power are also present around 12~c/d and 5~c/d. 
Their  distribution across the line profile is discontinuous, but patches of power corresponding to these 
periodicities are present all across the line profile and extend even towards the line wings.

The extraction of frequencies from the iterative least-squares fitting procedure was performed paying 
particular attention to the possibility that the strong daily aliases produced by the single-site sampling 
of our data, and the limited frequency resolution resulting from the observing baseline, can lead to a wrong 
identification of the true frequencies.
We computed spectra with an increasing number of known constituents and with various possible combinations of 
frequencies considering not only the highest peaks (which indeed could be the aliases of the true frequencies)
but also contiguous features at $\pm$ 1~c/d, following the procedure suggested by \citet{mantegazza95b}. 
The least-squares fitting procedure described above allowed us to extract iteratively the
following independent frequencies, reported in order of detection: 21.28~c/d, 5.06~c/d 12.02~c/d, 18.09~c/d, 
and 11.70~c/d. It is worth noting that the second
frequency extracted is very close to the frequency of the mode photometrically detected
by \citet{auvergne79}.

Fig.~\ref{figspettrof} shows the pixel-by-pixel power spectra and the corresponding global least-squares 
spectrum obtained at the end of the iterative procedure when all the known constituents have been removed 
from the data. Most part of the signal disappeared, but weak residual peaks at low frequencies are 
still present. A further iterative step of computation, however, did not allow us to identify firmly 
any additional periodic component. This is probably due to the limited data window of our data set. 
More observations on a longer baseline might contribute to clarify this point.

\begin{center}
\begin{figure}
\includegraphics[width=10cm]{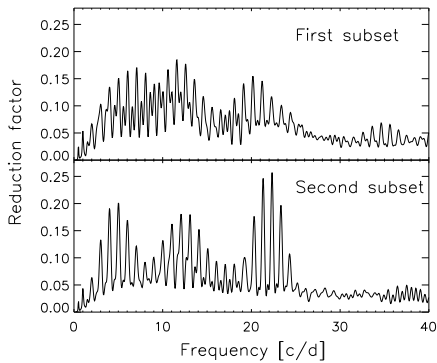}
\caption[]{Global least-squares power spectra without known constituents for the first (upper panel) 
and second subset (lower panel).}
\label{subsets}
\end{figure}
\end{center}

In order to investigate further the reliability of the 5-frequencies least-squares fit solution, 
we divided the data set in two subsets 3 nights long (see Table~\ref{log}), and analysed them separately,
with the aim of verifying if the same frequencies occur in both subsets. The power spectra without known 
constituents computed for the two subsets are shown in Fig.~\ref{subsets}; they appear quite noisy,
as expected, but the same three power concentrations detected in the whole data set spectrum are stil present in 
both subsets. The frequency resolutions of the two subsets are 0.22 and 0.23~c/d, respectively, 
to be compared with the 0.095~c/d for the complete data set.

Though the separation of data into two subsets is in principle essential for determining the reliability
of the identified frequencies, since in our case the data sampling in each subset is poor, only the gross features
can be taken into consideration. From an analysis of the power spectra of the two subsets it appears that the main
frequency at 21.28~c/d is present in both the subsets (with the aliases at the frequencies of 20.21~c/d and 22.30~c/d), 
as well as the one at 5.08~c/d (with the aliases at the frequencies of 5.00~c/d and 7.08~c/d), and probably 
the one at 12.02~c/d (but not present in the first subset); the frequencies at 11.70~c/d and 18.09~c/d appear only
when the whole set of data is considered.
From this analysis we can conclude that: i) the main frequency at 21.28~c/d is almost certainly present, but from the
present data alone it cannot be ruled out that the true frequency might actually be 20.28~c/d or 22.28~c/d; ii) 
the frequency at 5.06~c/d is probably present, but the true frequency might well be 4.06~c/d or 6.06~c/d; 
iii) the frequency at 12.02~c/d
might be present, but certainly not secure; iv) the frequencies at 11.70~c/d and 18.09~c/d are uncertain.

Although the spectral resolution of our data does not allow any rigourous mode identification, some 
useful clues for a very broad characterization of the detected frequencies can be derived from the 
analysis of the phase behaviour across the line profile for each solution of the final multiple-sinusoidal fit.
Actually it is well known that the phase diagrams give a full and independent representation of the 
line-profile variability caused by each individual pulsation mode and are considered useful tools for 
characterizing the detected variability (see \citet{telting97}).
Only the phase diagrams referring to the components  at 21.28~c/d, 18.09~c/d and 11.70~c/d show a 
clear signature (see Fig.~\ref{amp_phi_bis}), while those of the components at 5.06~c/d and 12.02~c/d 
are quite uncertain  and do not supply any useful discriminant. In  Fig.~\ref{amp_phi_bis} the 
wavelengths of the line profile have been converted into Doppler velocities assuming zero velocity 
at the centre of the line. The error bars correspond to the formal errors derived from the least-squares fit.
The phase diagrams of both the main component at frequency $21.28\,{\rm~c/d}$ and the component 
at  $18.09\,{\rm~c/d}$ show the typical behaviour of high-azimuthal order retrograde modes, i.e. their 
phase increases in the same direction of the stellar rotation, namely with the wavelength.
According to the commonly adopted expression $\exp[i(m \phi+2\pi\nu t)]$, which describes the  
time and longitude dependence of a pulsation at frequency $\nu$, it follows that both the cited 
modes are characterized by positive $m$-values.
The number of phase-lines in both the diagrams indicates that the azimuthal order should be quite hi\begin{table}
\caption{Observed oscillation frequencies and the present tentative at mode identification (see Section~3).}
\label{obs}
\begin{center}
\begin{tabular} {lcc}
\hline
\hline
Freq. (c/d)&  $l$ & $m$ \\
   &             &  \\
\hline
21.28 & $\geq10$  & $10, 11, 12$  \\
18.09 & $\geq8$ & 8, 9, 10  \\
12.02 & ? & ?  \\
11.70 & $\leq 4$ &  $-4 \leq m \leq -1 $ \\
5.06 & $\leq 4$& $-4 \leq m \leq 4$  \\
\hline
\end{tabular}
\end{center}
\end{table}gh 
for both terms. In order to get a quantitative estimate of the $m$-value for both modes we measured 
the shift in velocity $\delta v$ between the two central phase-lines measured at the centre of the 
line profile and deduced the separation $\delta\phi$ in longitude between the two equiphasic points 
from the relation: $\delta\phi = 2 {\rm arcsin} (\delta v/(2v_{\rm e}$)), where $v_{\rm e}$ is the 
equatorial rotation velocity, assuming for simplicity an equator-on view of the star $(i= 90^\circ)$. 
Since  $\vert m\vert = 360/\delta\phi$, we obtained $\vert m\vert  = 11.47 \pm 0.7$ for the 
component at 21.28~c/d and $\vert m \vert  = 9.16 \pm 0.7$ for the mode at 18.09~c/d. 
The derived $\vert m \vert$-value for the component at 21.28~c/d is in excellent agreement 
with that reported by \citet{kennelly92}.
If we assume again an equator-on geometry, the probability to be detected is higher 
for sectorial modes ($l = \vert m\vert$) since they are mainly focused at the equator
of the star. In this hypothesis the two components at 21.28~c/d and 18.09~c/d should be 
$l = m = 11$ and $l= m = 9$ retrograde modes, respectively. Other interpretations are obviously 
possible since it has been demonstrated that tesseral modes with $l -\vert m\vert$ small, observed at
an inclination angle of about 50 degrees, might produce similar moving patterns as sectorial modes 
observed equator-on (i.e. \citealt{schrijvers97}). The presence of retrograde modes in  $\delta$ Scuti stars, 
though much less frequent than prograde ones, is not uncommon (\citealt{mantegazza00}, \citealt{mantegazza01} 
and \citealt{kennelly98}). The phase diagram relevant to the term at frequency 11.70~c/d appears compatible 
with a low-degree prograde {$m$ $<0$) mode.

\begin{figure}
\includegraphics[width=10cm]{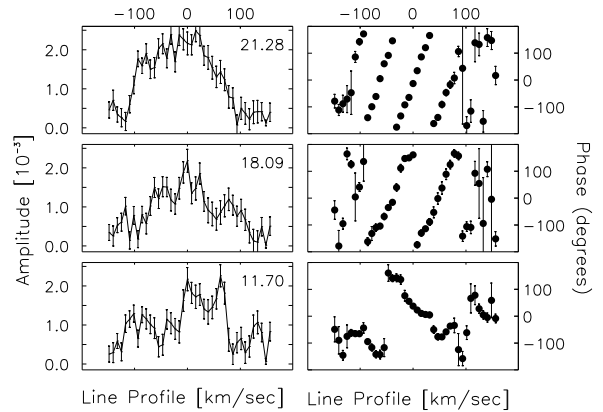}
\caption[]{Behaviour of amplitutes (left panels) and the phases (right panels) across the profile of 
the Ti{\sc ii} 4571.917{\AA} line of the periodic components 21.28, 18.09 and 11.70~c/d detected in the 
line-profile variation analysis. Wavelengths along the line profile have been converted into Doppler 
velocities relative to the line centre. The bars are the formal errors as derived by the global least-squares 
simultaneous fit. The  frequency in~c/d of each component is reported at the top of each panel.}
\label{amp_phi_bis}
\end{figure}

In order to catch additional information on the nature of the detected periodicites, we computed 
the equivalent width and the first three moments (which describe the centre-of-mass velocity of the line, 
the line width, and the line skewness, respectively) of the Ti{\sc ii} 4571.917{\AA} line profile,  
as a function of time, following the prescriptions given by \citet{balona96} and \citet{schrijvers97}.
Since the moments are quantities integrated over the whole line profile they are expected to be essentially 
sensitive to low-degree modes ($l \leq 4$). We then analyzed the resulting time series by adopting the software 
package for Fourier analysis and least-squares multiple-sinusoids
fitting, PERIOD04 \citep{lenz05} to search for periodicities. Useful results have been obtained 
only from the zero, second and third moments (see Fig.~\ref{momenta}), while the periodogram of the first 
moment is quite noisy.

\begin{figure}
\centering
\includegraphics[width=10cm]{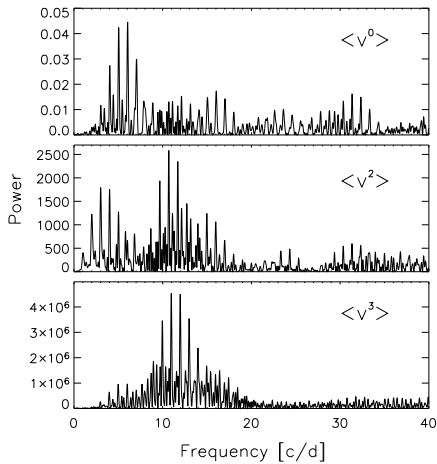}
\caption[]{Power spectra of the moments of zero  (upper panel), second (middle panel) and third order (bottom panel) 
obtained by the software package PERIOD04.}
\label{momenta}
\end{figure}

The periodogram of the equivalent width of the line, shows a concentration of power around the frequency 
6.05~c/d which could be the 1~c/d alias of the frequency 5.06~c/d, detected also in the line-profile 
variations and probably corresponding to the dominant photometric mode.
The second-order moment variability is characterized by two significant components at 10.69~c/d and 3.01~c/d, 
respectively. The former of the two frequencies, which appears to be slightly favoured by the analysis 
compared with the adjacent peak at 11.69~c/d, could actually be the one-day alias of the term at 11.70~c/d, 
also detected in the line-profile variation analysis.
The periodogram of the third moment shows a clump of power in the frequency range
$5 - 20\,\rm{c/d}$ with a significant feature present at 11.69~c/d. Other significant terms peak 
at 10.97~c/d, 17.43~c/d, and 5.08~c/d.
Once again the last value appears to be consistent with the result of the pixel-by-pixel line-profile 
variation analysis which gives 5.06~c/d.
The presence of the components at 5.06~c/d and 11.70~c/d in the variability of the first 
three moments might support their identification in terms of low $l$  ($\leq 4$) modes.
Moreover, the simultaneous presence of the term at 11.69~c/d
in the power spectra of both the second and third moments might allow us to rule out the  
possibility that the mode might be axisymmetric ($m=0$) (\citealt{balona86}).
As may be expected for high-degree pulsation modes, in the Fourier analysis of the moments 
there is no trace of neither the components at 21.28~c/d nor at 18.09~c/d, detected in the line-profile 
variation analysis and interpreted as high-azimuthal-order retrograde modes.

A summary of the results obtained is given in Table \ref{obs}
in which oscillation frequencies, tentative harmonic degree $l$, and azimuthal order $m$ of the detected modes are listed.

\section{Conclusion}
The new spectroscopic data of $\gamma$\,Bootis presented here confirm the early detection of the main 
pulsation component at 21.28\,~c/d found by \citet{kennelly92}.
Additional frequencies were detected at 18.09\,~c/d, 12.02\,~c/d,  11.70\,c/d and
5.06\,c/d.  The analysis of the phase diagrams relevant to each component
supplied useful hints for characterizing the detected variability. In particular,
the two terms at 21.28\,~c/d and  18.09\,~c/d appear to be  high-azimuthal order retrograde modes. 
In the hypothesis of an equator-on view of the star we suggest a tentative interpretation in terms of 
sectorial modes for both of them with $m = l = 11$ and $m = l = 9$, respectively. The identification 
proposed for the main component is in general good agreement with the conclusion of the work by \citet{kennelly92}. 
Moreover, the  component at 11.70\,~c/d is probably a low-$l$ prograde mode. The frequency analysis 
of the first three moments seems to confirm this interpretation and supplies
a constraint on $m$, suggesting that the mode is not axisymmetric.
The phase diagrams of the two additional terms at 5.06~c/d and 12.02~c/d are quite uncertain and 
do not  supply any useful discriminant.

The results obtained strongly encourage the planning of new observations with a higher S/N ratio 
and spectral resolution in order to get an unambiguous discrimination among the possible identifications  
of the modes by performing a direct fitting of the line-profile variations to pulsational models. Simultaneous 
photometric observations could also allow to model at the same time the flux variations in the line profiles.

Preliminary calculations of evolutionary models of $\gamma$\,Bootis constrained within the uncertainties
of the present parameter determinations indicate that this star has a mass of about $2{\rm M}_{\odot}$ and 
an age of about $0.9 {\rm Gyr}$, an immediate post-main sequence sub-giant star.

\begin{table}
\caption{Observed oscillation frequencies and the present tentative at mode identification (see Section~3).}
\label{obs}
\begin{center}
\begin{tabular} {lcc}
\hline
\hline
Freq. (c/d)&  $l$ & $m$ \\
   &             &  \\
\hline
21.28 & $\geq10$  & $10, 11, 12$  \\
18.09 & $\geq8$ & 8, 9, 10  \\
12.02 & ? & ?  \\
11.70 & $\leq 4$ &  $-4 \leq m \leq -1 $ \\
5.06 & $\leq 4$& $-4 \leq m \leq 4$  \\
\hline
\end{tabular}
\end{center}
\end{table}

\section*{Acknowledgments}
This work was supported by the European Helio- and Asteroseismology Network (HELAS), a major 
international collaboration funded by the European Commission's Sixth Framework Programme.
This research has made use of the SIMBAD database, operated at CDS, Strasburg, France.
This publication makes use of data products from the Two Micron All Sky Survey, which is a joint 
project of the University of Massachusetts and the Infrared Processing and Analysis Center California 
Institute of Technology, funded by the National Aeronautics and Space Administration and the National Science Foundation.
We are grateful to W. Dziembowski for useful discussions, G. Mignemi for his help during the observing run, 
and G. Ventimiglia for her contribution in the data analysis.

\end{document}